\documentclass[conference]{IEEEtran}
\usepackage{stmaryrd}
\usepackage{amsfonts}
\usepackage{amssymb}
\usepackage{stfloats}
\usepackage{cite}
\usepackage{graphicx}
\usepackage{psfrag}
\usepackage{subfigure}
\usepackage{amsmath}
\usepackage{array}
\usepackage{eurosym}
\usepackage{color}
\usepackage{amsthm}
\usepackage[utf8]{inputenc}
\usepackage[english]{babel}
\usepackage{algpseudocode}
\usepackage[hyperfootnotes=false]{hyperref}
\hypersetup{
 colorlinks=false,
 citecolor=Violet,
 linkcolor=Red,
 urlcolor=Blue}

\ifCLASSINFOpdf
\else
\fi
\begin{document}
%
\title{$\beta$-expansion: A Theoretical Framework for Fast and Recursive Construction of Polar Codes}
\newtheorem{Thm}{\bf Theorem}
\newtheorem{Lem}{\bf Lemma}
\newtheorem{Cor}{\bf Corollary}
\newtheorem{Def}{\bf Definition}
\newtheorem{Exam}{\bf Example}
\newtheorem{Alg}{\bf Algorithm}
\newtheorem{Prob}{\bf Problem}
\newtheorem{Rem}{\bf Remark}
\newtheorem{Prop}{\bf Proposition}



\author{
\IEEEauthorblockN{\hspace{-0.3cm}Gaoning He, Jean-Claude Belfiore, Ingmar Land, Ganghua Yang}
\IEEEauthorblockA{Paris Research Center, Huawei Technologies\\
20 quai du Point du Jour, Boulogne Billancourt, France}
\and
\IEEEauthorblockN{Xiaocheng Liu, Ying Chen, Rong Li, Jun Wang}
\IEEEauthorblockA{Hangzhou Research Center, Huawei Technologies\\
360 Jiangshu Road, Binjiang, Hangzhou, China}
\and
\IEEEauthorblockN{\hspace{5.5cm} Yiqun Ge, Ran Zhang, Wen Tong}
\IEEEauthorblockA{\hspace{5.5cm} Ottawa Research Center, Huawei Technologies\\
{\hspace{5.5cm} 303 Terry Fox Drive, Ottawa, Canada}}
}




%


\maketitle

\begin{abstract}

In this work, we introduce $\beta$-expansion, a notion borrowed from number theory, as a theoretical framework to study fast construction of polar codes based on a recursive structure of universal partial order (UPO) and polarization weight (PW) algorithm.
We show that polar codes can be recursively constructed from UPO by continuously solving several polynomial equations at each recursive step.
From these polynomial equations, we can extract an interval for $\beta$, such that ranking the synthetic channels through a closed-form $\beta$-expansion preserves the property of nested frozen sets, which is a desired feature for low-complex construction.
In an example of AWGN channels, we show that this interval for $\beta$ converges to a constant close to $1.1892 \approx 2^{1/4}$ when the code block-length trends to infinity.
Both asymptotic analysis and simulation results validate our theoretical claims.
\end{abstract}


%
\IEEEpeerreviewmaketitle

\section{Introduction}

Polar code, introduced by Arikan \cite{Arikan-09}, is the first code with an explicit construction to achieve the channel capacity for many classes of channels. 
The key idea of polar codes lies in the famous phenomenon of channel polarization, which transforms physical channels into virtual channels (or called synthetic channels).
These virtual channels tend to either extreme good (reliability equals to 1) or extreme bad (reliability equals to 0) when the block-length $N$ goes to infinity. In other words, channels are polarized.
In this case, the best strategy is to transmit useful information bits only in the positions of good channels and leave the positions of bad channels "frozen".
As a consequence, the question of how to efficiently identify the positions and rank the reliability order of these good and bad channels becomes critical for polar code construction.

Unfortunately, up to now, efficient construction only exists for binary erasure channel (BEC)~\cite{Arikan-09}. However, polar code is supposed to work at least for additive white Gaussian noise (AWGN) channels in the 5th generation wireless systems.
In the case of AWGN channels, several techniques are commonly used, for example, density evolution (DE)~\cite{DE-1} and Gaussian approximation (GA) of density evolution~\cite{GA-1}.
However, the main drawback of DE/GA is their high computational complexity, which scales linearly with the code block-length, and therefore unacceptable for practical systems with varying parameters such as block-length and code rate.
They are even infeasible to be used for an on-the-fly implementation of a low-latency encoder/decoder.

While the low-complex construction problem remains unsolved for the AWGN channel, it has been shown recently in \cite{Schurch-16} that a certain partial order relation exists between the reliabilities of the synthetic channels of a polar code.
This discovery may greatly reduce the complexity of polar code construction to a sublinear level, since these partial orders are deterministic and {\textit{universal}}, in the sense that it holds for any transmission channel.
It is further shown in \cite{Urbanke-17} that if we take the advantage of the partial order, we can construct a polar code of block-length $N$ by only computing the reliability of roughly a fraction $1/\log^{3/2}N$ of the synthetic channels.
Separately in~\cite{R1-167209}, a closed-form algorithm called polarization weight is proposed to characterize the reliability order of synthetic channels for AWGN channels with low computational complex.
These two independent works enlighten us to pursue a theoretical explanation towards a fast construction scheme of polar codes for practical transmission channels such as the AWGN channel.

In this paper, we analyse the properties of universal partial order (UPO) and discover that UPO has an inherent recursive structure.
We explore the theory of polarization weight algorithm by introducing a number theoretical concept called $\beta$-expansion.
From this theory and the recursive structure of UPO, we find that polar codes can actually be efficiently constructed from UPO by applying $\beta$-expansion at each recursive step with a carefully chosen $\beta$.

The paper is organized in the following form: We first revisit the concept of channel polarization and UPO of polar codes in section II. 
In section III, we show the recursive structure of UPO.
In section IV, we study $\beta$-expansion theory of polarization weight algorithm. 
We show in section V how $\beta$-expansion can help towards a fast construction.
Finally, we provide asymptotic analysis and numerical results in section V followed by conclusions in section VII\\

\section{Preliminaries}

\subsection{Channel polarization}

Channel polarization is a key phenomenon in polar codes. 
It consists in "splitting" the channel $W: \mathcal{X}\rightarrow \mathcal{Y}$ into a pair of channels $W^{0}: \mathcal{X}\rightarrow \mathcal{Y}^2$ and $W^{1}: \mathcal{X}\rightarrow \mathcal{X}\times \mathcal{Y}$, defined as
\begin{eqnarray}
\label{W0}
W^{0}(y_1,y_2|x_1) \!\!\!\!&=&\!\!\!\! \frac{1}{2}\sum_{x_2\in \mathcal{X}} W(y_1|x_1 \oplus x_2) W(y_2|x_2), \\
\label{W1}
W^{1}(y_1,y_2,x_1|x_2) \!\!\!\!&=&\!\!\!\! \frac{1}{2} W(y_1|x_1 \oplus x_2) W(y_2|x_2).
\end{eqnarray}
After applying $n$ times this "splitting" operation, from the original channel $W$, we obtain $N=2^n$ different channels $\lbrace W^0_N, W^1_N, \cdots,W^{N-1}_N\rbrace$, which are called "synthetic channels".

{\Def{(Synthetic channel): 
Let $( b_{n-1}, \ldots, b_1, b_0 )$, $b_k \in \lbrace 0, 1\rbrace$, be the binary expansion of integer $i\in [0, N-1]$ over $n$ bits with the most significant bit on the left. The synthetic channels are defined as
\begin{equation*}
W^{i}_N = ((W^{b_0})^{b_2}\cdots)^{b_{n-1}},\ i=0,\ldots, N-1
\end{equation*}
where $(\cdot)^{b_k}$ is obtained from equation \eqref{W0} and \eqref{W1}.}
}

{\Def{(Reliability measure):
If the mutual information $I(W^i_N)> I(W^j_N)$, or equivalently if the Bhattacharyya parameter \cite{Arikan-09} $Z(W^i_N)< Z(W^j_N)$, we say the synthetic channel $W^i_N$ is more reliable than $W^j_N$, denoted as $W^i_N \succ W^j_N$, or simply $i \succ j$.
$I(\cdot)$ and $Z(\cdot)$ are defined as
\begin{eqnarray*}
I(W)\!\!\!\!&=&\!\!\!\! I(X;Y), \\
Z(W)\!\!\!\!&=&\!\!\!\! \sum_{y\in\mathcal{Y}} \sqrt{p_{Y|X}(y|0) p_{Y|X}(y|1)}.
\end{eqnarray*}
where $p_{Y|X}(y|x)$ is the transmission probability.
}
}


\subsection{Universal partial order (UPO)}

In \cite{Schurch-16} it is proved that a partial order of reliability measure exists for any symmetric channels with binary inputs.
They are “partial” because they are insufficient to form a fully ordered sequence covering all $N$ bit positions, each corresponds to an index of the synthetic channel. 
It is further shown in \cite{Urbanke-17} that when having ranked all these indices, only a fraction $1/\log^{3/2}N$ of the synthetic channels remain to be ranked according to the channel.

Given any pair of synthetic channel indices $(x,y)$, their reliability relation may be determined by applying the following rules: 
\begin{itemize}
\item \textbf{Addition}: If a binary representation of the index of a synthetic channel is $(a,b,1,c)$, then it must be less reliable than the synthetic channel whose index has a binary representation $(a,b,0,c)$. 
Such "$1\succ 0$" pattern can happen on one bit or multiple bits. For example:
\begin{eqnarray*}
2\ (0,1,{\underline{0}}) &\prec & 3\ (0,1,{\underline{1}}) ,  \\
9\ (1,{\underline{0,0}},1)  &\prec & 15\ (1,{\underline{1,1}},1).
\end{eqnarray*}
where in this example $(1,0,0,1)$ is the binary representation of synthetic channel index $9$.

\item \textbf{Left-swap}: If a binary representation of the index of a synthetic channel is $(a,0,b,1,c)$, then it must be less reliable than the synthetic channel whose index has a binary representation $(a,1,b,0,c)$. Such "$0..1 \prec 1..0$" pattern can occur multiple times, and $0,1$ do not need to be adjacent to each other. For example:
\begin{eqnarray*}
2\ (\underline{0},\underline{1},0) &\prec & 4\ (\underline{1},\underline{0},0) ,  \\
12\ (\underline{0},1,\underline{1},0,0)  &\prec & 24\ (\underline{1},1,\underline{0},0,0).
\end{eqnarray*}
\end{itemize}
However, for some pair of synthetic channel indices $(x,y)$, their reliabilities cannot be determined by the rule of Addition or Left-swap or their combination. For example:
\begin{eqnarray*}
3\ (0, 1, 1) & \text{and} & 4\ (1, 0, 0) ,  \\
7\ (0, 1 ,1 ,1) & \text{and} & 12\ (1, 1, 0, 0) .
\end{eqnarray*}
We say that these orders are unknown to UPO.


\section{Properties of universal partial order}

{\Prop{ (Nested and Symmetric)
\label{Prop-Nested}
The universal partial order of polar codes has two important properties: 

\begin{itemize}
\item \textbf{Nested}: The orders determined in code length of $N$ remain unchanged in code length of $2N$.  
 
\item \textbf{Symmetric}: The order of $x\prec y$ and the order of $(N-1-x) \succ (N-1-y)$ are twin pairs for a polar code of length $N$.
\end{itemize}
}}

\begin{proof}
The Nested property is easy to be proved, since a binary expansion $(a,b,c)$ in length $N$ can be expressed as $(0,a,b,c)$ in length $2N$, which doesn't affect the rule of Addition and Left-swap.

To prove the Symmetric property, we let the binary expansion of $x$ be $(x_{n-1},...,x_1,x_0)$ and the binary expansion of $y$ be $(y_{n-1},...,y_1,y_0)$. Now assume $x\prec y$, we must have either for some $i\in [0,N-1]$
\begin{equation}
\label{Sym-Add}
x_i(=0) \prec y_i(=1)\ \ \text{(Addition)} 
\end{equation}
and/or for some $i,j\in[0,n-1], i\neq j$
\begin{equation}
\label{Sym-Swap}
x_i(=0),...,x_j(=1) \prec y_i(=1),...,y_j(=0)\ \ \text{(Left-swap)}.
\end{equation}
Now, let $\overline{x}$ be the one's complement of $x$. We write the binary expansion of $(N-1-x)$, the symmetric node of $x$, as $(\overline{x}_{n-1},...,\overline{x}_1,\overline{x}_0)$, and the binary expansion of $(N-1-y)$, the symmetric node of $y$, as $(\overline{y}_{n-1},...,\overline{y}_1,\overline{y}_0)$. 
From \eqref{Sym-Add} and \eqref{Sym-Swap}, we must have $(N-1-x)\succ (N-1-y)$. This is because we have either $\overline{x}_i(=1) \succ \overline{y}_i(=0)$ and/or $\overline{x}_i(=1),...,\overline{x}_j(=0) \succ \overline{y}_i(=0),...,\overline{y}_j(=1)$. 
This proves the symmetric property.
\end{proof}

Now, we are ready to show the recursive structure of the UPO. Let us first define $\mathbf{UPO}_n$ as the minimum set of UPO relation $x\prec y$ (denoted as $\{x,y\}$) of a polar code with length $N=2^n$, where $x,y$ are integers in $[0,N-1]$. Here, the minimum set means that if we have a relation chain $x\prec y \prec z$, we will express them as several sub-chains between only the most adjacent nodes, i.e., $x \prec y$ and $y \prec z$.
We have:
\begin{eqnarray*}
 \mathbf{UPO}_1 \!\!\! &=&\!\!\! \{\{0,1\}\}, \\
 \mathbf{UPO}_2 \!\!\! &=&\!\!\! \{\{0,1\},{\color{red}\{1,2\}},{\color{blue} \{2,3\}}\}, \\
 \mathbf{UPO}_3 \!\!\! &=&\!\!\! \{\{0,1\},\{1,2\},\{2,3\}, {\color{red}\{2,4\},\{3,5\}}, \\  && {\color{blue} \{4,5\},\{5,6\},\{6,7\}}\}, \\
 \mathbf{UPO}_4 \!\!\! &=&\!\!\! \{\{0,1\},\{1,2\},\{2,3\}, \{2,4\},\{3,5\}, \\  && \{4,5\},\{5,6\},\{6,7\} \\
 && {\color{red} \{4,8\},\{5,9\},\{6,10\},\{7,11\}, }\\
 && {\color{blue}\{8,9\},\{9,10\},\{10,11\}, \{10,12\},\{11,13\},} \\  && {\color{blue}\{12,13\},\{13,14\},\{14,15\}} \}
\end{eqnarray*}
where the orders in black color in level $n$ are inherited from UPO in the upper level $n-1$, i.e., $\mathbf{UPO}_{n-1}$ (corresponds to the nested property in Proposition~\ref{Prop-Nested}). 
The orders in blue color represent the symmetric part of the orders in black color (corresponds to the symmetric property in Proposition~\ref{Prop-Nested}), and the orders in red color represents the new orders when constructing $\mathbf{UPO}_n$ from $\mathbf{UPO}_{n-1}$.

\begin{figure*}
  \centering
  \includegraphics[scale=0.40]{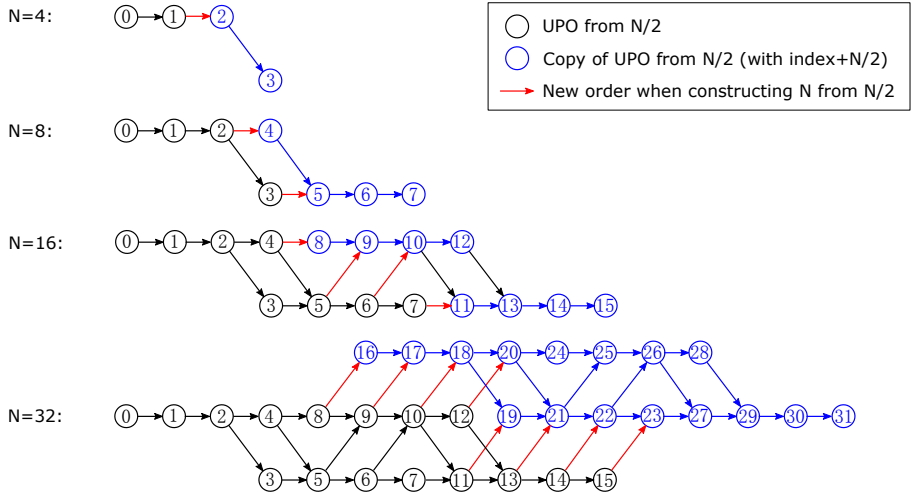}
  \caption{Illustration of recursive construction structure of universal partial orders of polar codes. The directed arrow between two nodes (synthetic channel indices) $x\rightarrow y$ indicate that the source node $x$ has lower reliability than the destination node $y$. }
\label{fig-UPO}
\end{figure*}

Generally speaking, any UPO of length $2^n$ ($\mathbf{UPO}_n$) can be recursively generated from $\mathbf{UPO}_1=\{\{0,1\}\}$ by applying the same rule.
This phenomenon is more recognizable from Figure~\ref{fig-UPO}.



\section{$\beta$-expansion: theory of PW algorithm}

The notion of polarization weight (PW) is first introduced in \cite{R1-167209}, where a closed-form algorithm is used to fully characterize the reliability orders of synthetic channels of a polar code.

{\Def{
\label{Def-PW}
(PW algorithm) 
Consider a synthetic channel index $x$ and its binary expansion $B=(b_{n-1},\ldots,b_1,b_0)$ over $n$ bits with the most significant bit on the left.
Its polarization weight is defined as
\begin{equation}
\label{Def-betaexpansion}
f^{\mathrm{PW}}: x \mapsto \sum^n_{i=1} b_i \beta^i
\end{equation}
}
where $\beta$ is a carefully chosen number. For example, $\beta=2^{1/4}$ is suggested in~\cite{R1-167209}.
}

In mathematics, this notion is called $\beta$-expansions~\cite{be-1,be-2} introduced by Re\'nyi as a number theoretical concept in 1957.
$\beta$-expansions are a generalization of decimal expansions and can be seen as a non-integral representation (with base $\beta$) of any real number $y$.

{\Exam{
Let $\beta=2^{1/4}$ and block-length $N=2^n=16$. The $\beta$-expansion of the synthetic channel with index $3$, i.e., $B_3=(0,0,1,1)$, can be computed as
\begin{equation*}
w_3= 0 \cdot 2^{3/4} + 0 \cdot 2^{2/4} + 1 \cdot 2^{1/4} + 1 \cdot 2^{0/4} = 2.189\cdots
\end{equation*}
Similarly, we can write the $\beta$-expansions of all the synthetic channels as
\begin{eqnarray*}
\mathbf{w} \!\!\!\!\!&=&\!\!\!\!\! \{0.000\ 1.000\ 1.189\ 2.189\ 1.414\ 2.414\ 2.603\ 3.603\\
\!\!\!\!\!&&\!\!\!\!   \ 1.682\ 2.682\ 2.871\ 3.871\ 3.096\ 4.096\ 4.285\ 5.285 \}
\end{eqnarray*}
where a larger value of $w_x$ indicates a higher reliability of the synthetic channel with index $x$. By sorting $\mathbf{w}$, we obtain a total order
\begin{eqnarray*}
\mathbf{Order}= \{  0\ 1\ 2\ 4\ 8\ 3\ 5\ 6\ 9\ 10\ 12\ 7\ 11\ 13\ 14\ 15 \}.
\end{eqnarray*}
}}

\begin{figure*}
  \centering
  \includegraphics[scale=0.9]{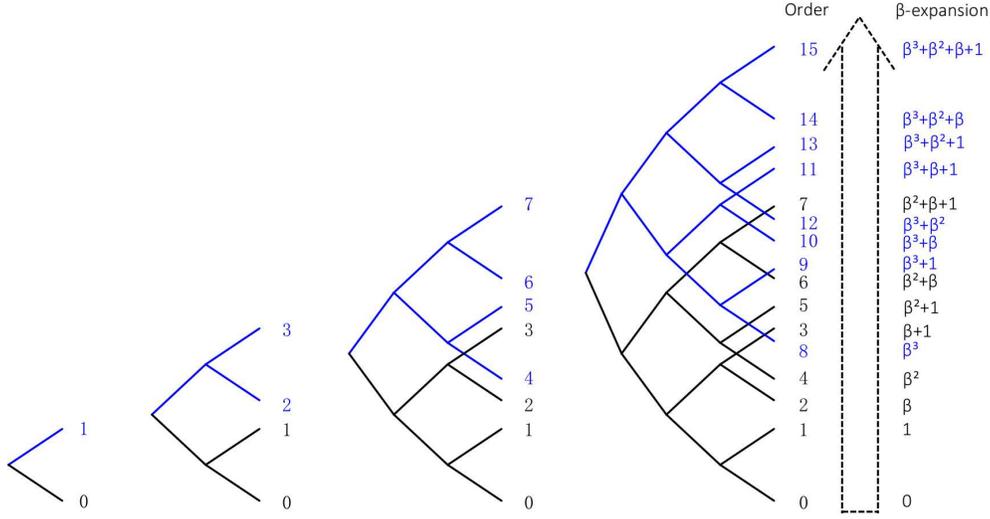}
  \caption{Illustration of nested structure of $\beta$-expansions with $\beta=2^{1/4}$ for blcok-length $N = 2, 4, 8, 16$.}
  \label{fig-nested}
\end{figure*}

The advantage of PW algorithm is that it provides a neat and low-complex method to fully rank the reliability of synthetic channels while keeping the property of nested frozen sets when the code length $2^n$ grows from $n=1$ to $+\infty$.

{\Prop{(Nested structure)
PW algorithm preserves the feature of nested code construction (or nested frozen sets) for polar codes. 
}
}
\begin{proof}
Similar to the proof of Proposition~\ref{Prop-Nested}, any arbitrary binary expansion $(a,b,c,\ldots)$ in length $N$ can be expressed as $(0,a,b,c,\cdots)$ in length $2N$. But this does not change the outcome of PW algorithm according to equation~\eqref{Def-betaexpansion}.
\end{proof}

Nested structure is a highly desired feature when designing code construction with variable block-lengths (which is the case for wireless systems). 
It helps to greatly reduce the overall construction complexity, in the sense that a construction at block-length $N$ can be kept and reused at longer block-length $2N, 4N,\cdots$. 
Figure~\ref{fig-nested} shows the nested structure of $\beta$-expansions with $\beta=2^{1/4}$ for block-length $N=2, 4, 8, 16$.

{\Prop{
When $\beta >1$, any sequence derived from PW algorithm respects the UPO of polar codes.
}}
\begin{proof}
First, the Addition rule is respected by $\beta$-expansion, because $0<\beta^j$ holds for any non-negative integer $j$.
Second, the left-swap rule is also respected, since when $\beta > 1$ the inequality $\beta^j< \beta^{j+k}$ holds for any non-negative integer $j$ and positive integer $k$. This completes the proof.
\end{proof}

From this proposition, we know at least the sequence derived from PW algorithm does not violate the orders from UPO as long as $\beta\in(1,+\infty)$.
This also implies that if $\beta$ picks a specific number in $(1,+\infty)$, PW algorithm introduces new orders which do not belong to UPO.
Now, the question is which value $\beta$ should take such that PW algorithm or $\beta$-expansions can give a good sequence.

It is known that $\beta$-expansions are not necessarily unique. A counter-example is when $\beta = \phi=\frac{1+\sqrt{5}}{2}$, the famous golden ratio, we have $\phi + 1 = \phi^2$.
This means that $\beta$-expansions may become ill-conditioned (in the sense that it loses the ability to fully rank the sequence) when $\beta$ equals to some algebraic numbers that are real number roots of 0-1 polynomials (whose exponents are either $0$ or $1$ or $-1$). 
For example, $\phi$ is the root of polynomial $x^2-x-1=0$.
This is the reason we pointed out in Definition~\ref{Def-PW} that the value of $\beta$ shall be carefully chosen.
To address this issue, we have the following theorem:


{\Thm{
For a polar code of length $2^n (n\geq 3)$, there exists a set of ascending numbers $\mathcal{A}_n=\{a_0, a_1, a_2,\cdots\, a_{i_{\max}},+\infty \}$, where $a_0=1$ and $a_1, a_2,\ldots, a_{i_{\max}}$ are algebraic numbers, such that for any $\beta$ falls in the intervals of $(a_i,a_{i+1}), \forall i\geq 0$, $\beta$-expansion is unique and results in a different ordered sequence.
Moreover, the set $\mathcal{A}_n$ is nested, i.e., $\mathcal{A}_n \subset \mathcal{A}_{n+1}$.
}}

\begin{proof}
From equation \eqref{Def-betaexpansion}, we can analyse the $\beta$-expansions for $n=1$ to $4$, as follows


\begin{itemize}

\item For $n=1$ and $2$, the sequence order is unique. This is because we have $\beta>1$ by default.
\item For $n=3$, there is only one undecided relation, i.e., $( \beta+1, \beta^2 )$, whose order depends on the value of $\beta$. 
In order to distinguish their order, we need to solve the polynomial equation $x^2-x-1=0$, and we have \begin{equation*}
\mathcal{A}_3=\{1,1.618,+\infty\}.
\end{equation*}

When $\beta \in (1, 1.618)$, we have $\beta+1 > \beta$ , the corresponding sequence is
\begin{equation*}
0\rightarrow 1\rightarrow 2 \rightarrow {\color{red}4 \rightarrow 3} \rightarrow 5 \rightarrow 6\rightarrow 7.
\end{equation*}
When $\beta \in (1.618, +\infty)$, we have $\beta+1 < \beta$, the corresponding sequence is
\begin{equation*}
0\rightarrow 1 \rightarrow 2 \rightarrow {\color{red} 3 \rightarrow 4} \rightarrow 5 \rightarrow 6 \rightarrow 7.
\end{equation*}
That is, when $\beta$ falls in any interval of $\mathcal{A}_3$, $\beta$-expansion is unique and gives a different sequence.

\item For $n=4$, there are $4$ undecided relations, including the previous one $( \beta+1, \beta^2 )$ and two of its equivalent forms, i.e.,
\begin{eqnarray*}
(\beta+1, \beta^2),\ (\beta^2+\beta, \beta^3),\ (\beta^2+\beta, \beta^3)\\
(\beta+1, \beta^3),\ (\beta^2+1, \beta^3),\ (\beta^2+\beta+1, \beta^3)
\end{eqnarray*}
In order to fully distinguish their orders, we need to solve $4$ polynomial equations
\begin{eqnarray}
x^3-x^2-x-1&=&0  \label{eq:poly1}\\
x^3-x^2-1&=&0 \label{eq:poly3}\\
x^2-x-1&=&0 \label{eq:poly2}\\
x^3-x-1&=&0 \label{eq:poly4}
\end{eqnarray}
and we have 
\begin{equation*}
\mathcal{A}_4=\{1,1.325, 1.466, 1.618, 1.839, +\infty\},
\end{equation*}
where $1.325, 1.466, 1.618, 1.839$ are the roots of the four polynomial equations~\eqref{eq:poly1}-\eqref{eq:poly4}, respectively.
\end{itemize}

\begin{table}[h!]
\centering
\caption{$\beta$-expansions for $n=1$ to $4$}
\begin{tabular}{|c|l|}
\hline
$n$ & {$\beta$-expansions} \\
\hline \hline
$1$ & \small{$0,\ 1$} \\
\hline
$2$ & \small{$0,\ 1,\ \beta,\ \beta\!\!+\!\!1$} \\
\hline
$3$ & \small{$0,\ 1,\ \beta,\ \beta\!\!+\!\!1,\ \beta^2,\ \beta^2\!\!+\!\!1,\ \beta^2\!\!+\!\!\beta,\ \beta^2\!\!+\!\!\beta\!\!+\!\!1$} \\
\hline
 & \small{$0,\ 1,\ \beta,\ \beta\!\!+\!\!1,\ \beta^2,\ \beta^2\!\!+\!\!1,\ \beta^2\!\!+\!\!\beta,\ \beta^2\!\!+\!\!\beta\!\!+\!\!1,$} \\
$4$ & \small{$\beta^3,\ \beta^3\!\!+\!\!1,\ \beta^3\!\!+\!\!\beta,\ \beta^3\!\!+\!\!\beta+1,\ \beta^3\!\!+\!\!\beta^2,\ \beta^3\!\!+\!\!\beta^2+1,$} \\
 &  \small{$\beta^3\!\!+\!\!\beta^2\!\!+\!\!\beta,\ \beta^3\!\!+\!\!\beta^2\!\!+\!\!\beta\!\!+\!\!1,$} \\
\hline
\end{tabular}
\end{table}

Similarly, we can iteratively run this process for the case of $n>4$.
It is apparent that the set $\mathcal{A}_n$ is nested, because of the nested property of $\beta$-expansion. More precisely, if we have a undecided relation $(x,y)$ in the sequence of length $N$, it must remain in the sequence of length $2N$, which results in a nested set for the corresponding algebraic number.
This completes the proof.
\end{proof}

\section{An example of fast construction for AWGN channels}

As discussed previously, UPO is a fundamental basis of polar code construction, which can determine many of the reliability orders of the synthetic channels, however, it cannot fully rank the whole sequence.
On the other hand, although techniques such as DE/GA are able to rank the whole sequence, they are not suitable for practical implementation due to their high complexity.
To compensate the drawbacks of both approaches, in this section, we present a fast construction method based on UPO and $\beta$-expansions.

We first give some intuition through the following Figure~\ref{fig-betaexpansion16}. 

\begin{figure}[!h]
  \centering
  \includegraphics[scale=0.35]{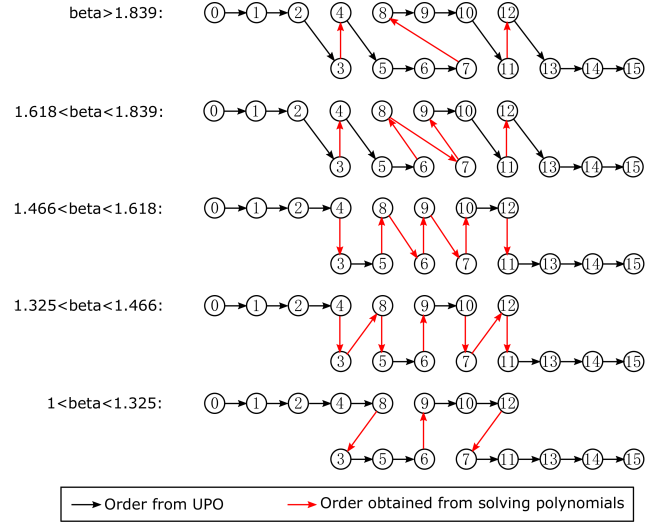}
  \caption{$\beta$-expansions for a polar code of block-length $N=16$}
  \label{fig-betaexpansion16}
\end{figure}

We observe that when $\beta$ falls in different interval in the set $\mathcal{A}_4=\{1,1.325, 1.466, 1.618, 1.839, +\infty\}$, $\beta$-expansion gives in a different ordered sequence for polar codes of block-length $N=16$.

From DE/GA, we know that for any SNR the sequence order for AWGN channels is
\begin{eqnarray*}
0 \rightarrow 1 \rightarrow 2 \rightarrow 4 \rightarrow 8 \rightarrow 3 \rightarrow 5 \rightarrow 6 \rightarrow 9 \\
 \rightarrow 10 \rightarrow 12 \rightarrow 7 \rightarrow 11 \rightarrow 13 \rightarrow 14  \rightarrow 15 
\end{eqnarray*}
which corresponds to the interval of $\beta \in (1,1.325)$.
This means if we carefully choose a interval for $\beta$, we can control the outcome of $\beta$-expansion so that it gives the same or approximately the same good sequence compared with DE/GA.

In the following example, we show how this idea works for AWGN channels and how the interval of $\beta$ behaves when the block-length grows from $8$ to $1024$.
\begin{itemize}

\item {$\mathbf{N=8}$}

Thanks to the recursive structure of UPO, we can construct an ordered sequence of length $N=8$ from two sequences of length $N=4$, as shown in Figure~\ref{fig:4to8}.
Although the order of $(3,4)$ is unknown to UPO, from DE-GA we know $4 \prec 3$ for any SNR.
This decides the interval $(1, 1.618)$ for $\beta$.
\begin{figure}[!h]
  \centering
  \includegraphics[scale=0.40]{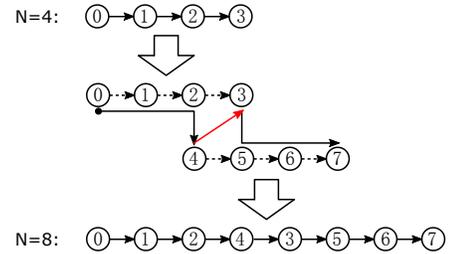}
  \caption{Illustration of sequence construction from $N=4$ to $8$.}
  \label{fig:4to8}
\end{figure}

\item {$\mathbf{N=16}$}

Similar to N=8, we can construct the sequence of length $N=16$ from two sequences of length $N=8$, as shown in Figure!\ref{fig:8to16}.
\begin{figure*}
  \centering
  \includegraphics[scale=0.45]{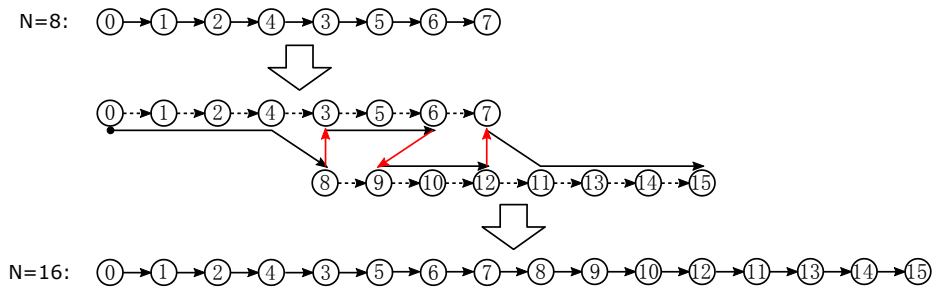}
  \caption{Illustration of sequence construction from $N=8$ to $16$.}
  \label{fig:8to16}
\end{figure*}
We have 3 new pairs $(6,9), (3,8)$ and $(7,12)$ that are unknown to UPO. Again, from DE/GA we know $6 \prec 9, 8 \prec 3$ and $12 \prec 7$ for any SNR.
Accordingly, the range of $\beta$ is reduced from $(1, 1.618)$ to $(1,1.325)$.

\item {$\mathbf{N=32}$}

We have 5 new pairs that are unknown to UPO: $(28,15), (14,19), (24,13), (24,11)$ and $(24,7)$. 
Again, from DE-GA we have
\begin{eqnarray*}
28 \prec 15,\ 24 \prec 11 &\Longleftrightarrow & \beta<1.221 \\
14 \prec 19 &\Longleftrightarrow & \beta<1.325 \\
24 \prec 13 &\Longleftrightarrow & \beta<1.272
\end{eqnarray*}
The order $(24,7)$ is in general SNR-dependent for AWGN channel.
However, a greedy simulations tell us that the order of $7 \prec 24$ has an overall better performance than that of $7 \succ 24$ for nearly all SNR values. Therefore, we choose
\begin{eqnarray*}
7 \prec 24 &\Longleftrightarrow & \beta>1.179
\end{eqnarray*}
Accordingly, the range of $\beta$ is reduced from  to $(1,1.325)$ to $\in (1.179,1.221)$.

\item {$\mathbf{N>32}$}

We can repeat this searching process and we have the following table. We see that the interval of $\beta$ converges to a constant close to $1.1892\approx 2^{1/4}$, and the number of new pairs is less than $20\%$ of the total number of synthetic channels.
\begin{table}[h!]
\centering
\caption{Convergence of interval for $\beta$ and the number of new pairs when generating sequence from length $N$ to $2N$, for AWGN channels}
\begin{tabular}{|l|l|l|}
\hline 
$N \rightarrow 2N$ & Interval of $\beta$ & Number of new pairs  \\
\hline \hline
$16 \rightarrow 32$ & $(1.179, 1.221)$ & $5$ \\
\hline
$32 \rightarrow 64$ & $(1.179, 1.194)$ & $10$ \\
\hline
$64 \rightarrow 128$ & $(1.185, 1.190)$ & $\sim 30$ \\
\hline
$128 \rightarrow 256$ & $(1.1885, 1.190)$ & $\sim 50$\\
\hline
$256 \rightarrow 512$ & $(1.18875, 1.18952)$ & $\sim 90$ \\
\hline
$512 \rightarrow 1024$ & $(1.189, 1.18932)$ & $\sim 200$ \\
\hline
\end{tabular}
\label{Table:covergenceofbeta}
\end{table}
\end{itemize}

\section{Asymptotic analysis and numerical results}

In this section, we study the asymptotic behaviour of $\beta$-expansion when $n\rightarrow \infty$.
In fact, the problem of studying distributions of the $\beta$-expansion can be transformed to another concept called Bernoulli convolution \cite{Bernoulli-1, Bernoulli-2}. Bernoulli convolution defines a probability measure
\begin{equation*}
\nu_\lambda = \sum_{i=1}^{+\infty} b_i \beta^{i}
\end{equation*}
while assuming that $B = (\ldots, b_1, b_0)$ is a sequence of independent binary random variables satisfying

\begin{equation*}
\mathbf{Pr}\lbrace b_i=1 \rbrace=\mathbf{Pr} \lbrace b_i=0 \rbrace = 0.5,\ \forall i
\end{equation*}

It has been proved in~\cite{Bernoulli-1} that the measure $\nu_\lambda$ is either {\textit{absolutely continuous}} or {\textit{singular}}. 
Clearly, we need the property of absolutely continuous for $\nu_\lambda$ to be able to fully rank the reliability of infinite number of synthetic channels. 

It is further proved in \cite{Bernoulli-2} that if $\beta$ is of the form of $2^{1/k}$, where $k$ is any positive integer, then $\nu_\lambda$ is absolutely continuous.
This result justifies that the choice of $\beta=2^{1/4}$ in \cite{R1-167209} satisfies the necessary condition of absolutely continuous.
When $k=4$, we have $\beta=1.1892$, which coincides with our observation in Table~\ref{Table:covergenceofbeta}.

\begin{figure}
  \includegraphics[width=9cm]{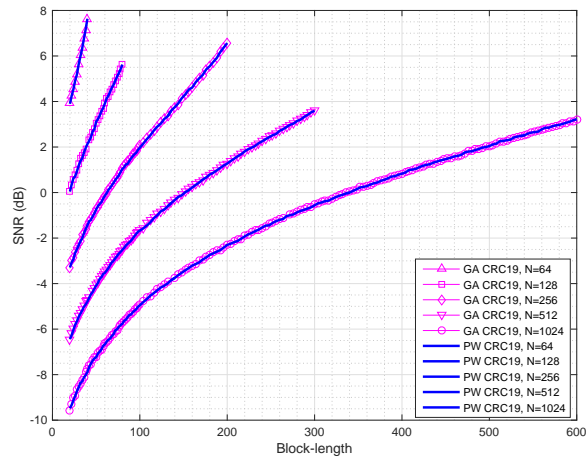}
  \caption{Comparison of PW algorithm with $\beta=1.1892$ and Gaussian approximation for different block-length $N$ (code length in number of code bits). Simulation assumptions are: BLER $= 0.001$, QPSK modulation, AWGN channel, successive cancellation list (SCL) decoder with list size $8$ and extra 19-bit CRC.}
  \label{fig:betaGA}
\end{figure}

In Figure~\ref{fig:betaGA}, we compare the performance of GA and low-complex $\beta$-expansion construction with $\beta=1.1892$ for AWGN channels.
The target block error rate (BLER) is 0.001 and  modulation QPSK is used.
The decoder is successive cancellation list (SCL) decoder with list size $8$ and extra 19-bit CRC.
As expected, $\beta$-expansion performs equally well as GA but with much lower complexity.

\section{Conclusion}
In this work, we consider the problem of fast construction of polar codes. 
We develop a theoretical framework based on $\beta$-expansion to explain the principle of polarization weight algorithm.
We not only show the condition of uniqueness of $\beta$-expansion, but also provide the link between intervals of $\beta$ values and different sequence ordering.
We discover a recursive structure of universal partial orders, from which we propose a fast and recursive method to construct polar codes by solving polynomial equations in each recursive step.
We show in the example of AWGN channels that the interval for $\beta$ converges towards a constant close to $1.1892$ when block-length tends to infinity.
Finally, both asymptotic analysis and numerical results confirm that the proposed low-complex method has the same performance compared with GA.






%

\end{document}